\documentclass[final,5p,times,twocolumn,letterpaper]{elsarticle}

\usepackage{amsthm}         
\usepackage{amsmath} 	   
\usepackage{amssymb}       
\usepackage{amsfonts}
\usepackage{graphicx} 	    
\usepackage{verbatim}
\usepackage{color}
\usepackage{colortbl}
\usepackage{float}
\usepackage{multirow}

\usepackage{amsthm}         
\usepackage{amsmath} 	   
\usepackage{amssymb}       
\usepackage{amsfonts}
\usepackage{mathtools}
\usepackage{graphicx} 	    
\usepackage{verbatim}
\usepackage{epsfig,bbm}
\usepackage{color}
\usepackage{colortbl}
\usepackage{float}
\usepackage{multirow}
\usepackage{lipsum}
\usepackage{tensor}
\usepackage{enumitem}

\usepackage{subcaption}

\definecolor{babyblue}{rgb}{0.54, 0.81, 0.94}
\definecolor{corn}{rgb}{0.98, 0.93, 0.36}

\begin{document}

\begin{frontmatter}

\title{Initial conditions problem in cosmological inflation revisited
}

\author[add0]{David Garfinkle} 
\author[add1]{Anna Ijjas} 
\author[add2]{Paul J. Steinhardt
}

\address[add0]{Department of Physics, Oakland University, Rochester, MI 48309, USA}

\address[add1]{Center for Cosmology and Particle Physics, Department of Physics, New York University, New York, NY 10003, USA}
\address[add2]{Department of Physics, Princeton University, Princeton, NJ 08544, USA}


\date{\today}

\begin{abstract}
We present first results from a novel numerical relativity code based on a tetrad formulation of the Einstein-scalar field equations combined with recently introduced gauge/frame invariant diagnostics indicating that inflation does not solve the homogeneity and isotropy problem beginning from generic  initial conditions following a big bang.
\end{abstract}


\end{frontmatter}

\section{Introduction}

The prime theoretical challenge for any complete cosmological theory is to explain the extraordinary homogeneity, isotropy and flatness of the early expanding universe and its  remarkably low gravitational entropy.  All the successes of the $\Lambda$CDM picture, which describes the evolution of the universe beginning from cosmic nucleosynthesis (temperatures below $\sim 10$~MeV or so),  rely on there having been some preceding mechanism that smoothed and flattened spacetime first.  Furthermore, any proposal claiming to explain the origin of the cosmic microwave background (CMB) temperature fluctuations as arising from quantum fluctuations is based on a perturbation calculation that presumes a spacetime that is nearly flat Friedmann-Robertson-Walker (FRW) and has low gravitational entropy.  Since these conditions are highly non-generic in general relativity and certainly not expected following a big bang, a complete cosmological theory must include a dynamical mechanism that transforms generic 
 initial conditions to non-generic low-entropy flat FRW conditions.   

Inflation was first introduced as a classical dynamical mechanism capable of achieving this goal \cite{Guth:1980zm,Linde:1981mu,Albrecht:1982wi}.  Furthermore, presuming inflation succeeded, theorists next considered what would happen to quantum fluctuations if there was an additional 60 or more $e$-folds of inflation after smoothing and flattening \cite{Bardeen:1983qw,PhysRevLett.49.1110,Hawking:1982cz,Starobinsky:1982ee}.  
Their calculations suggested that quantum fluctuations generated on scales much smaller than the Hubble radius could be stretched to super-Hubble scale wavelengths to form a spectrum of nearly scale-invariant adiabatic fluctuations that would source CMB temperature fluctuations and the formation of large-scale structure.  

 But does inflation achieve the goal it was originally designed to accomplish?   An affirmative answer would mean that, given the essential ingredients needed to have inflation -- an inflaton field $\phi$ with some scalar field potential $V(\phi)$ -- homogeneity, isotropy, flatness and low gravitational entropy would be probable outcomes for {\it generic} initial conditions.   There are several well-know reasons to be concerned whether this is true for inflation.  
  For example, inflation can only occur if the inflaton field happens to lie in 
  in a specific limited range of $V(\phi)$ that is far above the ground state and whose slope is sufficiently small; and only if the inflaton kinetic energy density is small compared to its potential energy density, even though small kinetic energies are not expected emerging from a big bang.   There is also the 
   quantum instability problem that leads to eternal inflation 
 and the conversion of spacetime into an infinite multiverse of volumes spanning all combinations of inhomogeneity, anisotropy and curvature, 
  \cite{Steinhardt:1982kg,Vilenkin:1983xq,Linde:1983gd}, providing no explanation why  our observed universe has the special properties it does.

In this paper, though, we put aside these concerns to focus on yet a broader issue: ignoring quantum effects and assuming gravity is described by the classical Einstein equations, does inflation smooth and flatten the universe beginning from the generic spacetime initial conditions expected when the universe emerges from a big bang -- the original motivation for introducing inflation?   The investigation is made possible by two recent theoretical developments:
\begin{itemize}
\item A numerical relativity code based on a modified tetrad formulation of the Einstein-scalar field equations which, unlike previous numerical relativity approaches to analyzing inflation, makes it possible to follow the evolution of spacetime as the inflaton field travels the entire range from the `flat' portion of the potential down to the potential minimum $V=0$ ({\it i.e.}, the entire duration of inflation).  
\item The identification of gauge/frame invariant diagnostics based on the Weyl curvature tensor, as recently introduced in Ref.~\cite{Ijjas:2023bhh}, that make it possible to quantitatively evaluate the genericity of the initial conditions and  the success or failure in reaching sufficiently smooth and flat final conditions.
\end{itemize}

\section{Numerical scheme}
\label{sec:formalism}
We numerically solve the (3+1)-dimensional Einstein-scalar field equations in mean-curvature-normalized, orthonormal tetrad form, as has been successfully 
implemented in early studies of contracting spacetimes \cite{Garfinkle:2008ei,Cook:2020oaj,Ijjas:2020dws,Ijjas:2021gkf,Ijjas:2021wml,Ijjas:2021zyf,Kist:2022mew}.   The numerical scheme is explained in great detail in \ref{app:numericalscheme}.   For simplicity, we present results in which the spatial variations are along a single spatial direction $x$; extension to variations along two and three dimensions will be presented in follow-up papers.  

The novel feature of the tetrad formulation for inflation compared to the studies of contracting spacetimes is that the coordinate time $t$ runs from zero to $+\infty$ (rather than from zero to $-\infty$) and is rescaled according to 
\begin{equation}
\label{newtime2}
\frac{{\rm d}\ln K}{{\rm d}t} =  -\mu(t),
\end{equation}
where $K\equiv K_a^a$ is the mean curvature,
\begin{equation}
\label{def-C02}
\mu(t) \equiv \frac{1}{\tilde{{\cal N}}_{\rm max}(t)},
\end{equation}
and $\tilde{{\cal N}}_{\rm max}(t)$ is the maximum value of the  coordinate lapse $N$ at time $t$ rescaled by the mean curvature divided by $\mu$, $\tilde{\cal N} \equiv {\textstyle \frac13}KN/\mu$ to become dimensionless. 
As we show in \ref{app:newtime}, our time coordinate $t$ then measures the maximum number of $e$-folds of inflation taking place in the course of the simulation.  The modification compared to simulations of contracting spacetimes (where we set $\mu(t)= 1$) is essential because we are using constant mean curvature time-slicing.  In the case of slow contraction, the mean curvature (proportional to $H$ in the homogeneous limit) changes by an exponential amount; but in the expanding case, it only changes by less than an order of magnitude throughout inflation, so a modified time-slicing is needed to be able to follow $e$-fold by $e$-fold for a sufficiently long time. 

As is common in numerical relativity to date,
we set the spatial metric of the initial $t_0$-hypersurface to be  conformally-flat,  
$\gamma_{ij}(t_0,\vec{x}) = \psi^4(t_0,\vec{x})\delta_{ij}$,
where $\psi$ denotes the conformal factor and use the York method \cite{York:1971hw} to specify constraint satisfying initial conditions. The components of the spatial 3-curvature tensor $\bar{n}_{ab}, \bar{A}_b$ and the tetrad vector components ${\bar{E}{}_a}^i$ are then fixed as follows: 
\begin{alignat}{2}
\label{eq3}
&\bar{n}_{ab}(t_0,\vec{x}) &&= 0 ,  \\
&\bar{A}_b(t_0,\vec{x}) &&= -2\psi^{-1}{\bar{E}{}_b}^i\partial_i\psi,\label{eq4} \\
& {\bar{E}{}_a}^i(t_0,\vec{x}) &&= \psi^{-2}(K_0/3)^{-1}{\delta_a}^i.
\end{alignat}
where 
$\bar{A}_b \equiv {\textstyle \frac12} \epsilon_{b}{}^{cd}\bar{N}_{cd}$
is the antisymmetric part of $\bar{N}_{ab}$ and $\bar{n}_{ab} \equiv \bar{N}_{ab} - \epsilon_{ab}{}^c \bar{A}_c$ is the symmetric part.   N.B.  a `bar' on top of any variable henceforth corresponds to rescaling by the mean curvature ({\it i.e.}, dividing by $K/3$).

One  is then left with the freedom to specify the initial value of the mean curvature $K_0$, the initial scalar field distribution $\phi(t_0,\vec{x})$, the conformally  re-scaled initial scalar field velocity 
$\bar{Q}$,
as well as the divergence-free part of the conformally-rescaled shear tensor, $\bar{Z}_{a b}^{0} (t_0,\vec{x})\equiv \psi^6\bar{\Sigma}^0_{ab}$.

To explore a broad range of initial conditions, we specify the free data as follows:
\begin{equation}
\tensor*{\bar{Z}}{_i_j^0} = \left(\frac{K_0}{3}\right)^{-1}\times
\renewcommand*{\arraystretch}{1.3}
\begin{pmatrix}
b_2 &  \xi & 0  \\
\xi &   b_1+  a_1 \cos{x} &  a_2 \cos{x}  \\
0 
&    a_2 \cos{x}  &   -b_1 - b_2 - a_1 \cos{x} \end{pmatrix}
\label{ZZ2}
\end{equation}
where $a_1, a_2, b_1, b_2$ and $\xi$ are constants; 
\begin{align}
\begin{split}
\label{QQ2}
\bar{Q} &=\left({\textstyle \frac{K_0}{3}}\right)^{-1}\times \Big(q_{x} \cos{(m_x x + d_x)} + Q_0 \Big) \\
\phi &= f_x \cos{(n_x x + h_x)} + \phi_0,
\end{split}
\end{align}
where $Q_0, \phi_0, q_x,  f_x,  m_x,  n_x,  d_x$ and  $h_x$ are constant and denote the mean value, the amplitude, the mode number and the phase of the initial  velocity and field distribution, respectively. The choice of cosine reflects the fact that, for the numerical simulation, we choose periodic boundary conditions $0\leq x\leq2\pi n$ with $0$ and $2\pi n$ identified, where $n\geq1$ is an integer.
(Even though we only show a single mode for the shear and  the field's velocity, they can be replaced by a sum of different fourier modes with different amplitudes, wavenumbers and phases.)


The initial data is then completed by numerically computing the conformal factor, $\psi(t_0,\vec{x})$, and the longitudinal (non-vacuum) part of the shear tensor, $Z_{ab} - Z^0_{ab}$, using the Hamiltonian and momentum constraints:
\begin{alignat}{2}
\label{hamiltonian-constraint2}
& \partial_i\partial^i\psi 
&&+ {\textstyle \frac18}\psi^{-7} \left({\textstyle \frac{K_0}{3}}\right)^2\left(\tensor{\bar{Z}}{^i^j}\tensor{\bar{Z}}{_i_j} + \bar{Q}^2 \right) - {\textstyle \frac{1}{12}}\psi^5 \left(K_0^2 -  3V \right)   \\
& &&+ {\textstyle \frac18}\psi \partial^i\phi\partial_i\phi  = 0 ,
\nonumber\\
\label{momentum-constraint2}
& \partial^i\tensor{\bar{Z}}{_i_j}  &&= \bar{Q}\partial_j\phi.
\end{alignat} 
This approach enables us to freely vary all the physical degrees of freedom given the Hamiltonian and momentum constraints, which is essential for our goal of exploring the outcomes for generic initial conditions.  

Most previous numerical relativity studies of inflation have imposed additional constraints on the available degrees of freedom in order to simplify or avoid numerically solving the Hamiltonian and momentum constraint equations above.  For example, by choosing conditions with $\bar{Z}_{ij}=0$ and $\dot{\phi}=0$, the second term of the Hamiltonian constraint is eliminated.  In so doing,  the study is limited to a subset of initial conditions of measure zero.  Most significantly, these restricted choices artificially favor inflation, giving the misleading impression that inflation has no initial conditions problem.  We will return to this point and make it more quantitative using gauge/frame independent diagnostics.

We evolve the hyperbolic-elliptic system of partial differential equations by discretizing the equations using second order accurate spatial derivatives and a three-step method for time integration where we employ the Iterated Crank-Nicolson algorithm. At each sub-step, we
first solve the elliptic equation~\eqref{lapse-eq-Hmu} and then update the hyperbolic equations~(\ref{eq-E-ai-Hmu}-\ref{eq-S-a-Hmu}) to the next iterated Crank-Nicolson sub-step.

\section{Gauge/frame invariant diagnostics}

Recently, a set of diagnostics based on the Weyl tensor has been identified for evaluating the results of numerical relativity studies of cosmological spacetimes in a way that does not depend on the specific choice of formulation, coordinate gauge or frame~\cite{Ijjas:2023bhh}.  The diagnostics prove to be powerful quantitive tools for analyzing the genericity of initial conditions and the success or failure of a smoothing mechanism in reaching a flat FRW outcome. 

The {\it conformal Weyl curvature tensor} (or Weyl tensor, for short)  is the  trace-free part of the Riemann curvature tensor $\tensor{R}{_\mu_\nu_\sigma_\rho}$:
\begin{equation}
\label{Weyl-coo}
\tensor{\cal C}{_\mu_\nu_\rho_\sigma}
\equiv \tensor{R}{_\mu_\nu_\rho_\sigma}
+ {\textstyle \frac12}\left( \tensor{g}{_\mu_\nu_\rho_\zeta}\tensor{R}{^\zeta_\sigma} 
+ \tensor{g}{_\mu_\nu_\zeta_\sigma}\tensor{R}{^\zeta_\rho}\right)
 + {\textstyle \frac16}R \tensor{g}{_\mu_\nu_\rho_\sigma},
\end{equation} 
where $g_{\mu\nu}$ denotes the spacetime metric, $\tensor{R}{_\mu_\nu}\equiv \tensor{R}{^\sigma_\mu_\sigma_\nu}$ denotes the Ricci curvature tensor, $R\equiv \tensor{R}{^\mu_\mu}$ denotes the Ricci curvature scalar 
and $\tensor{g}{_\mu_\nu_\rho_\sigma} \equiv  g_{\mu\rho}g_{\nu\sigma} - g_{\mu\sigma}g_{\nu\rho}$; see \cite{Weyl:1918pdp}.
Here spacetime indices $(0-3)$ are Greek and spatial indices $(1-3)$ are Latin.  

The Weyl tensor describes the `non-local' part of the gravitational field, {\it i.e.}, inhomogeneities and anisotropies of the spacetime geometry that are not sourced by a local stress-energy source.
In (3+1)-dimensions, the Weyl tensor is generically non-zero and equal to zero  if and only if spacetime is conformally-flat.  

The (left) dual of the Weyl tensor is
\begin{equation}
\label{def-dualWeyl}
{}^{\ast}\tensor{\cal C}{_\mu_\nu_\rho_\sigma}
\equiv \frac12 \tensor{\chi}{_\mu_\nu^\tau^\zeta}\tensor{\cal C}{_\tau_\zeta_\rho_\sigma},
\end{equation}
with $ \tensor{\chi}{_\mu_\nu_\tau_\zeta}\equiv - \sqrt{|-g|}\tensor{\epsilon}{_\mu_\nu_\tau_\zeta}$ being the totally anti-symmetric Levi-Civita 4-form and $\tensor{\epsilon}{_\mu_\nu_\tau_\zeta}$ being the Levi-Civita tensor.

The Weyl tensor and its dual can be combined to yield two curvature invariants, which are the gauge/frame diagnostic quantities we want to evaluate in analyzing simulations:  (a) the {\it Weyl curvature}
\begin{equation}
\label{def-weylscalar}
{\cal C} \equiv \tensor{\cal C}{^\mu^\nu^\rho^\sigma}\tensor{\cal C}{_\mu_\nu_\rho_\sigma};
\end{equation}
 and, (b)  the {\it Chern-Pontryagin invariant}
\begin{equation}
\label{def-CPscalar}
{\cal P} \equiv  {}^{\ast}\tensor{\cal C}{^\mu^\nu^\rho^\sigma}\tensor{\cal C}{_\mu_\nu_\rho_\sigma}.
\end{equation}
 A flat FRW spacetime has Weyl tensor equal to zero and, hence, ${\cal C}=0$ and ${\cal P}=0$; conversely, a spacetime with non-zero  ${\cal C}$
and ${\cal P}$ is not flat FRW.

Penrose's Weyl Curvature Hypothesis~\cite{Penrose:1979azm} posited that the spacetime emerging from a big bang must have zero Weyl curvature tensor (or, equivalently, zero ${\cal C}$  and ${\cal P}$ ) in order to evolve into the homogeneous, isotropic and flat universe we observe today and ensure the gravitational entropy was negligible, as required for cosmological evolution consistent with the second law of thermodynamics.  However, the expectation is that the Weyl curvature is large after a bang due to the period of large quantum gravity fluctuations and strong coupling of all degrees of freedom that accompanies the bang.  Hence, Penrose's Hypothesis turns into a Puzzle:  what mechanism could cause an initial state with large Weyl Curvature to transform into a final state with vanishingly small Weyl Curvature?   Here is where numerical relativity becomes a critically important tool for cosmology -- as a means of evaluating dynamical mechanisms.

\section{Results} 

The numerical relativity litmus test for any smoothing and flattening mechanism, inflation or otherwise, is to show that generic initial states with {\it large} $\bar{\cal C}$ and $\bar{\cal P}$ evolve to final states with {\it sufficiently small} $\bar{\cal C}$ and $\bar{\cal P}$, where the bar means relative to the mean curvature.  Quantitatively, {\it large} means the rescaled ratios are order unity or greater.  {\it Sufficiently small} for cosmological purposes means 
 $\bar{\cal C} \sim {\cal O}( ({}^3\bar{R})^2)$ and $\bar{\cal P} \sim {\cal O}( ({}^3\bar{R})^2)$ should become much less than the quantum contributions to curvature fluctuations $({}^3\bar{R})^2$, which are known to be ${\cal O}(10^{-10})$ based on observations of the CMB; we will refer to this threshold as the {\it effectively flat FRW} condition. 

In this section, we present the litmus test results for the  inflationary potential $V(\phi) = (1/2) m^2 \phi^2$, a standard example of `large field' inflation, where, for simplicity, we solve the (3+1)-dimensional Einstein-scalar field equations assuming deviations from flat FRW along a single spatial dimension.  (Previous studies suggest that large field inflation is more robust to initial gradients than small field inflation \cite{Clough:2016ymm}, and so a good choice for our purpose.   In subsequent papers, we will show the results for a wider range of potentials and with deviations along two or three spatial dimensions.)  We have run hundreds of examples with similar results, but here we will only show one representative example with initial conditions
\begin{align*}
&K_0 = -3.00;m =3.8 \times 10^{-3};\\
&a_1=7,\; a_2=10,\; b_1=18,\; b_2=-1.5,\; \xi=0.01; \\
& f_x = 7.5,\; n_x=3,\; h_x =-1.1,\; \phi_0=12.5; \\
& q_x =2.5,\; m_x=4,\; d_x=-1.6,\; Q_0 =-5.
\end{align*}   
where $\phi$  is in units of reduced Planck mass ($8 \pi G=1$ where $G$ is Newton's constant) and  $m$ and spatial coordinates are in units of the initial value of $|K/3|$.    The time coordinate $t$ is related to $K$ according to Eq.~(\ref{newtime2}).  In the limit of homogeneous flat FRW initial conditions (fixing $\phi_0 = max[\phi(x)]=20$ for all $x$), $-K/3$ is the initial Hubble parameter, $t$ is $\ln \, a(t)$, the number of $e$-folds of expansion, and there are $ t \approx 85$ $e$-folds of inflationary expansion.   Because the study reported here only allows spatial variations along a single direction $x$, it suffices to present results for any physical quantity at time $t$ as a function of $x$.  The `box size' is $\Delta x = 2 \pi n/ |K/3|$ where $n\ge 1$ is an integer; results are qualitatively similar as $n$ is increased.  We use a box size with $n = {\cal O}(10)$ in order to keep track of the initial volume as $|K/3|$ decreases during the simulation.

\begin{figure}[!t]%
    \centering
\includegraphics[width=0.95\linewidth]{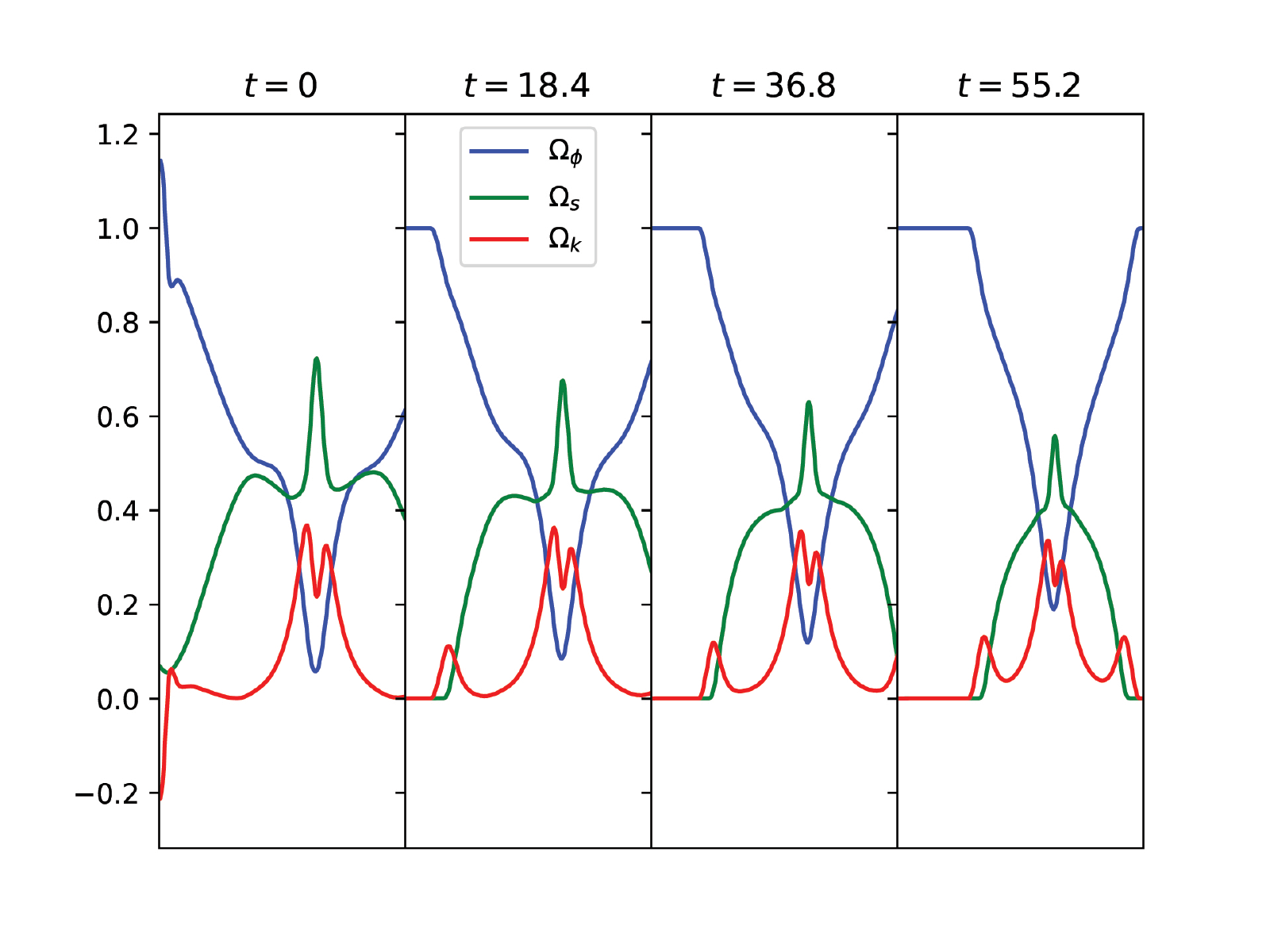}%
    \caption{Snapshots of the shear ($\Omega_s$), spatial curvature ($\Omega_k$) and scalar field energy density ($\Omega_{\phi}$) relative to the mean curvature as a function of time.  Two-thirds of the spatial range where smoothing is not observed begin with generic ${\cal O}(1)$ initial values of  the invariants $\bar{{\cal C}}$ and $\bar{{\cal P}}$ , as shown in  Figs.~\ref{fig2} and~\ref{fig3}. 
    In contrast, on the left third, the initial $\bar{{\cal C}}$ and $\bar{{\cal P}}$  have small values, a non-generic condition emerging from a big bang.}
    \label{fig1}%
\end{figure}

\begin{figure}[!t]%
    \centering
\includegraphics[width=0.95\linewidth]{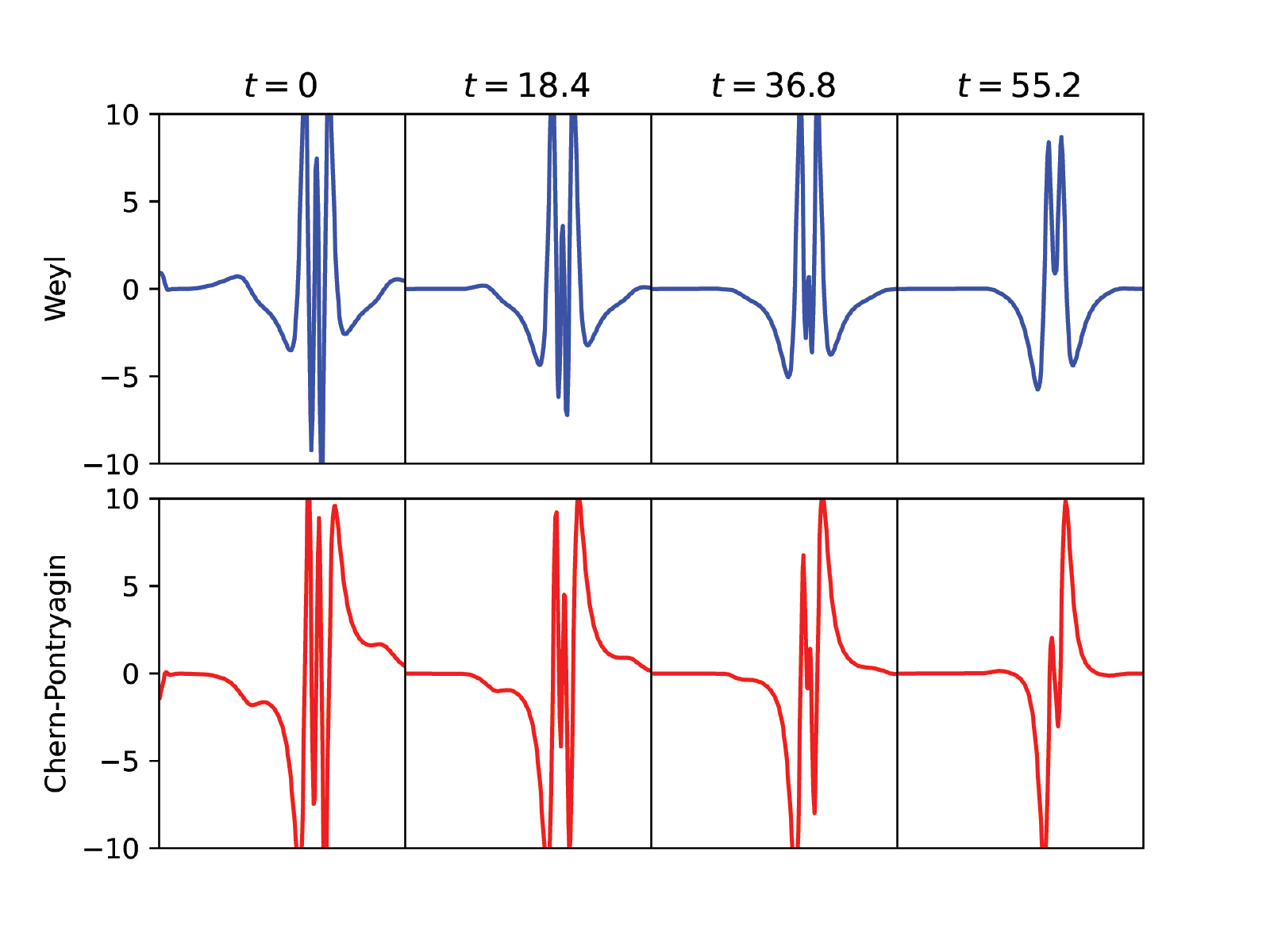}%
    \caption{The same sequence of snapshots as in Fig.~\ref{fig1} showing  $\bar{{\cal C}}$ and $\bar{{\cal P}}$. 
    The snapshots show that, except for the  left side, the initial values of  $\bar{{\cal C}}$ and $\bar{{\cal P}}$  are ${\cal O}(1)$ or more and the values at later times are much greater than $10^{-10}$, which signifies failing to reach flat FRW. }
    \label{fig2}%
\end{figure}

In order to properly interpret the results, we need to point out certain aspects of numerical relativity simulations that favor inflation.  (Notably, there is no analogy for simulations of slow contraction where we can define initial data that disfavors slow contraction everywhere in the simulation box \cite{Ijjas:2022qsv}.)  These aspects relate to constraints on spatial curvature, which is the strongest suppressor of inflation; for example, in a homogeneous  expanding universe, spatial curvature decays more slowly than the anisotropy.  
First, the convention of using a conformally flat spatial metric in applying the York method to select constraint-preserving initial conditions means that 
$\bar{n}_{ab}=0$, as in Eq.~\eqref{eq3} and $\bar{A}_b$ takes the form in Eq.~(\ref{eq4}) on the initial time slice.
 Deviations from zero spatial curvature are generated in subsequent evolution steps by the non-linear interaction terms in the Einstein equations.  However, because the simulations require periodic boundary conditions, regions of positive curvature must be accompanied  by regions with negative curvature and a region of negligible curvature in between.  Then, because an inflaton potential is necessarily positive, the total energy density of the scalar field ($\Omega_{\phi}$ when mean curvature normalized) is positive definite; consequently, regions with large negative curvature  $\Omega_k$ tend to be accompanied by a large and dominant $\Omega_{\phi}$, a combination that  has non-generic anomalously small  $\bar{{\cal C}}$ and $\bar{{\cal P}}$ and scalar field conditions that favor inflation.  All this has nothing to do with real physical cosmology; it is an artifact of using a conformally flat spatial metric at $t=0$  and periodic boundary conditions.  (In principle, more general metrics can be used with the York method but doing so in practice is highly non-trivial and has not yet been 
 developed for cosmology.)

This apparent disadvantage is actually an advantage, though, since it means that, in a single simulation, we can track using the same code regions with generic initial conditions  characterized by substantial $\bar{{\cal C}}$ and $\bar{{\cal P}}$ and regions with non-generic initial conditions.  In particular, we can observe directly how differently the two evolve and the degree to which inflation can only take hold if there are special initial conditions following a big bang.

Fig.~\ref{fig1} is a series of snapshots showing the evolution of $\Omega_{\phi}(x)$, $\Omega_k(x)$ and the mean curvature normalized shear $\Omega_s(x)$; the three must sum to unity according to the Hamiltonian constraint.   Fig.~\ref{fig2} shows snapshots at the same times of the  rescaled Weyl Curvature
 $\bar{{\cal C}}$ and Chern-Pontryagin invariant $\bar{{\cal P}}$.   The time $t$ is proportional to the number of $e$-folds of expansion in regions that are homogeneous and expanding, as explained in \ref{app:newtime}.

\begin{figure}[!t]%
    \centering
\includegraphics[width=0.95\linewidth]{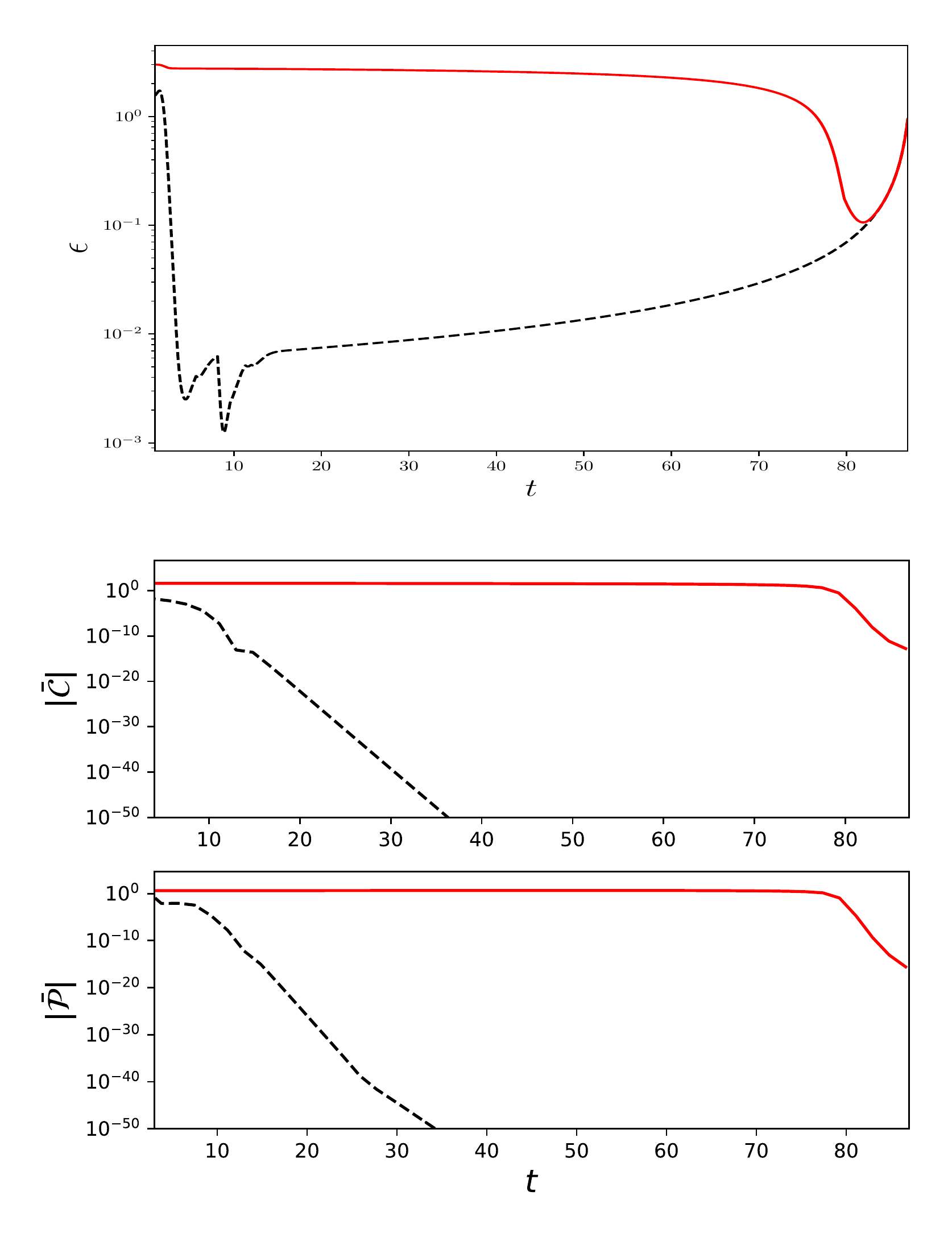}%
    \caption{The top panel shows $\varepsilon $ vs. $t$.   The red solid curve corresponds to a generic region of  spacetime in Fig.~\ref{fig1} where  the initial  $\bar{{\cal C}}$ and $\bar{{\cal P}}$ are  ${\cal O}(1)$ or greater.  Note that $\varepsilon >1$ over almost the entire evolution (inflation is suppressed).  The lower two panels show that  $\bar{{\cal C}}$ and $\bar{{\cal P}}$ remain larger than $10^{-10}$ during the entire evolution of the inflaton down the potential.  The black dashed curves show the  corresponding evolution for a highly non-generic point ({\it e.g.}, left side of the panels in Fig.~\ref{fig1}) where the initial $\bar{{\cal C}}$ and $\bar{{\cal P}}$ are set to small values and $\Omega_m$ is set to dominate right from the beginning.  }
    \label{fig3}%
\end{figure}

In the $t=0$ panels, a varying mix of comparable $\Omega_m$, $\Omega_s$ and $\Omega_k$ can be seen on the right two-thirds in Fig.~\ref{fig1}, accompanied by ${\cal O}(1)$ variations in $\bar{{\cal C}}$ and $\bar{{\cal P}}$ in Fig.~\ref{fig2};  the latter correspond to generic initial conditions emerging from the big bang.   These conditions remain as the inflaton field evolves down the `flat' ($V'/V \ll 1$ ) slow-roll portion of its potential.  Inflation is not occurring because of the non-linear interactions in the Einstein equations that prevent the shear and curvature from being quenched.  
By contrast, the far left hand side begins with non-generic conditions: large dominant $\Omega_m$ and small $\bar{{\cal C}}$ and $\bar{{\cal P}}$.  By the second panel, $t=18.4$ panel, $\Omega_s$, $\Omega_k$, $\bar{{\cal C}}$ and $\bar{{\cal P}}$ have all become negligible on the scale shown.

Fig.~\ref{fig3} provides  more quantitative detail.  The upper panel shows $\varepsilon  \equiv 1/\tilde{\cal N}$, where $\tilde{\cal N}$ is the rescaled lapse.  In the homogeneous limit, $\varepsilon$  is  the equation of state ({\it i.e.}, $\varepsilon \rightarrow \frac{3}{2}(1 + p/\rho)$, where $p$ is the pressure and $\rho$ is the energy density, as described in \ref{app:homogeneous}.  Slow-roll inflation down the potential corresponds to $\varepsilon \ll 1$ and monotonically increasing.

The red solid curves track $\varepsilon(t)$, $\bar{{\cal C}}$ and $\bar{{\cal P}}$, respectively, in the three panels at a point on the right two-thirds of the panels in Figs.~\ref{fig1} and~\ref{fig2} which begin with generic initial conditions; that is,  $\bar{{\cal C}}$ and $\bar{{\cal P}}$  are ${\cal O}(1)$ or greater.  
Clearly  $\varepsilon$ remains greater than one (no inflation) across nearly the entire period the inflaton evolves down its potential.  Near the end, $\varepsilon$ falls below than one, but there is little time left. The red solid curves in the  lower two panels of Fig.~\ref{fig3} show that, even at the end, $\bar{{\cal C}}$ and $\bar{{\cal P}}$  
do not reach the $10^{-10}$ threshold required to be effectively flat FRW.  
This illustrates that regions with generic initial conditions emerging from a big bang  are not smoothed enough to match the homogeneity and isotropy observed in the cosmic microwave background.   

Actually, the situation is worse than depicted so far.  It is not sufficient that $\bar{{\cal C}}$ and $\bar{{\cal P}}$  fall below  $10^{-10}$ by the end of the run. If inflation is to explain the  nearly scale-invariant spectrum of cosmic microwave fluctuations as originating from Bunch-Davies quantum fluctuations, the threshold must be reached with at least 60 $e$-folds of inflation remaining.  By this reckoning, {\it regions with generic initial conditions emerging from a big bang fail by a wide margin}.

For comparison, the black dashed curves follow $\varepsilon(t)$, $\bar{{\cal C}}$ and $\bar{{\cal P}}$ respectively, at  a point on the left side of the panels in Figs.~\ref{fig1} and~\ref{fig2} with non-generic initial conditions.  Here one can see at $t=18.4$ that $\varepsilon$ satisfies the requisite conditions, confirming that inflation has begun and continues for more than 60 $e$-folds before inflation ends (at $t \approx 85$).  The lower panels show that $\bar{{\cal C}}$ and $\bar{{\cal P}}$ begin small and decrease exponentially fast as a function of $t$.  Already at $t=18.4$, they have fallen below the threshold for effectively flat FRW.

\begin{figure}[!t]%
    \centering
\includegraphics[width=0.9\linewidth]{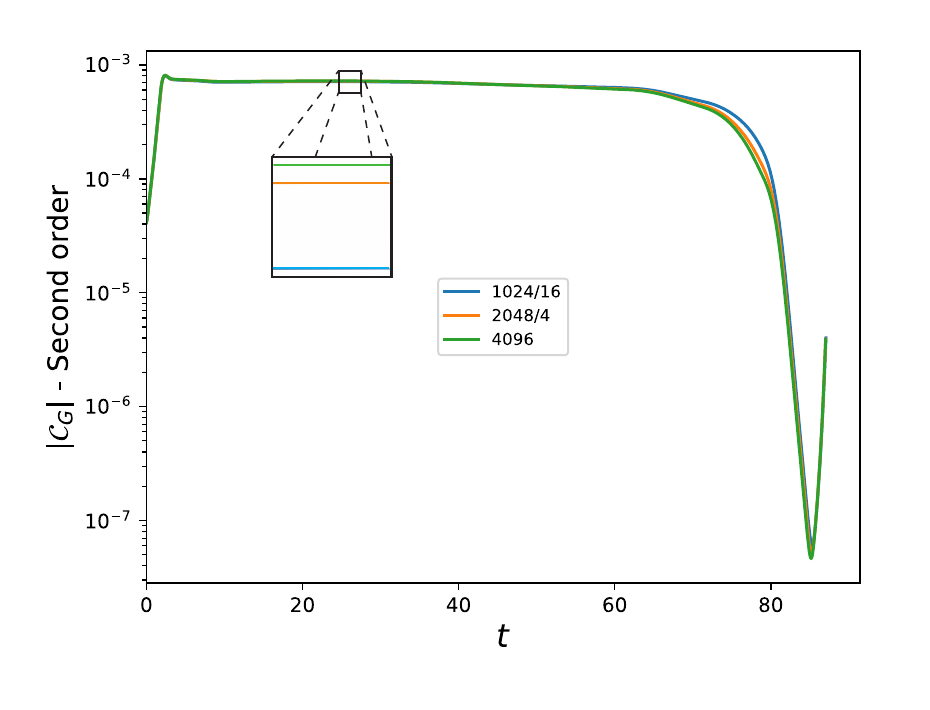}%
    \caption{The L2 norm of the Hamiltonian constraint $||{\cal C}_G||$  integrated over the spatial domain as a function of time.  Second order convergence requires that runs at three different spatial resolutions (here 1024, 2048 and 4096 gridpoints) differing by factors of two nearly overlap when the L2 norm is rescaled by powers of four; furthermore, as shown in the blowup, the rescaled curves should become closer as the resolution increases. }%
    \label{fig4}%
\end{figure}

As a check on our simulation results, we performed a standard set of numerical relativity convergence tests comparing runs at three different spatial resolutions for the same input parameters and computing the L2 norm of the constraints that must be satisfied.  As an example, we show in Fig.~\ref{fig4} the successful test of the Hamiltonian constraint $||{\cal C}_G||$ integrated over the spatial domain as a function of time. Runs at three different spatial resolutions differing by factors of two nearly overlap when the L2 norm is rescaled by powers of four; furthermore, as shown in the blowup, the rescaled curves become closer as the resolution increases.

\section{Discussion}
The numerical relativity results presented here have two advantages over previous numerical studies of inflation.  First, by introducing a modified tetrad formulation of the inflationary Einstein-scalar field equations that includes solving for the conformal factor $\psi$ in  Eq.~(\ref{hamiltonian-constraint2}) and the elliptic equation for the lapse in Eq.~(\ref{lapse-eq-Hmu}), we are able to freely specify all the physical degrees of freedom that remain after physical constraints (such as the Hamiltonian and momentum constraints) are satisfied and maintain a stable evolution code --  which is essential for exploring generic initial conditions.  

Some previous studies have tried to avoid numerically solving the Hamiltonian constraint  for the conformal factor by introducing approximate guesses,  whose errors can propagate through the entire evolution and cannot be rigorously accounted for; or 
 by choosing special initial conditions  that trivially satisfy Eq.~(\ref{hamiltonian-constraint2}) algebraically.   The problem with the second approach is that 
 it restricts the initial conditions to special choices of  zero measure, such as   ${\cal P}=0$ and $\dot{\phi}=0$.  These restricted choices
 strongly favor inflation, giving the misleading impression that inflation can generically take hold after a big bang.

  Second, invoking the gauge/frame invariant Weyl Curvature $\bar{{\cal C}}$ and Chern-Pontryagin invariant $\bar{{\cal P}}$ (normalized by the mean curvature), we are able to check in an objective quantitative way whether a given set of initial conditions is generic or not.  As Penrose has emphasized, the Weyl curvature tensor should be large (in comparison to the mean curvature) following a big bang, which corresponds to   $\bar{{\cal C}}$ and $\bar{{\cal P}}$ both being ${\cal O}(1)$.
 Furthermore, we have shown that  the same two invariants can be used as metrics to determine if a proposed smoothing mechanism
 fails to bring
spacetime sufficiently close to flat FRW to satisfy observational constraints on homogeneity, isotropy and flatness and with enough smoothing time remaining to generate a spectrum of nearly scale-invariant density perturbations.  For this purpose, it is important that, unlike some of the other formulations that have been tried, the tetrad code enables us to follow the evolution for the entire period the inflaton passes down its potential.   Finally, the invariants inform us whether the Weyl Curvature tensor, which Penrose explains as a measure of gravitational entropy~\cite{Penrose:1988mg}, fails to become  vanishing small as desired.
  
Our results for the numerous studies thus far indicate that inflation does not solve the homogeneity and isotropy problems   or satisfy the Weyl Curvature Hypothesis beginning from generic initial conditions following a big bang.  Our next steps are to expand our studies to other potentials and to include initial spatial variations along two and three spatial dimensions.   
 \\

\noindent
{\it Acknowledgments.} We thank Frans Pretorius for useful discussions.  The work of A.I. is supported by the
Simons Foundation grant number 947319. P.J.S. is supported in part by the
DOE grant number DEFG02-91ER40671 and by the Simons Foundation grant
number 654561.

\bibliographystyle{apsrev}
\bibliography{inflation}

\appendix

\section{Einstein-scalar field equations w/rescaled time coordinate in mean-curvature-normalized, orthonormal tetrad form }
\label{app:numericalscheme}

In order to keep the paper self-contained, we sketch out the derivation of the Einstein-scalar field equations in mean-curvature-normalized, orthonormal tetrad form using the rescaled time coordinate we introduced in Eq.~\eqref{newtime2}. For a detailed derivation, see, {\it e.g.}, \cite{Buchman:2003sq,Garfinkle:2008ei,Ijjas:2020dws}.

\subsection{Variables}

In orthornormal tetrad form, spacetime points are represented through four unit basis 4-vectors $\{\tensor{\bf e}{_\alpha^\mu}\}$ (as opposed to four coordinates) that form a local Lorentz frame with the local metric being (flat) Minkowski, {\it i.e.,} ${\bf e}_{\alpha}\cdot{\bf e}_{\beta}={\rm diag}(-1,1,1,1)$ and $\cdot$ denotes the inner product of the tetrads.
The Ricci rotation coefficients $\gamma_{\alpha\beta\lambda}={\bf e}_{\alpha}\cdot\nabla_{\lambda}{\bf e}_{\beta}$ define how the tetrad deforms when moving from one spacetime point to another. Here,  $\nabla_{\lambda}$ is the covariant derivative projected onto ${\bf e}_{\lambda}$.

The forty geometric variables of the formulation are given by the sixteen tetrad vector components  $\{\tensor{e}{_\alpha^\mu}\}$ and the twenty-four Ricci rotation coefficients, 
\begin{alignat}{2}
&\tensor{b}{_a} &&= \tensor{\gamma}{_a_0_0},\\
&\tensor{\Omega}{_a} &&={\textstyle \frac12} \tensor{\epsilon}{_a^b^c}\tensor{\gamma}{_b_c_0},\\
&\tensor{K}{_a_b} &&= -\tensor{\gamma}{_0_b_a},\\
&\tensor{N}{_a_b} &&= {\textstyle \frac12}\tensor{\epsilon}{_b^c^d}\tensor{\gamma}{_c_d_a},
\end{alignat} 
where $b_a$ and $\Omega_c$ are gauge variables, denoting the acceleration and the angular velocity of the spatial triad $\{{\bf e}_a,{\bf e}_b,{\bf e}_c\}$ relative to Fermi-propagated axes. The dynamical variables $K_{ab}$ and $N_{ab}$ denote the components of the shear and the spatial curvature tensor, respectively.

\subsection{Fixing the tetrad frame}
\label{app:gauge}

As in Refs.~\cite{Ijjas:2020dws,Ijjas:2021gkf,Ijjas:2021wml}, our tetrad frame of choice $\{{\bf e}_{\alpha}\}$ has 
{\it Fermi-propagated} axes (no non-physical rotations, {\it i.e.}, $\Omega_a\equiv0$) and the timelike congruence is {\it hypersurface orthogonal}  ({\it i.e.}, ${\bf e}_0$ is the future-directed unit normal to the spacelike hypersurfaces ${\cal T}_t$ of constant time $t$ and the spatial triad is tangent to $\{{\cal T}_t\}$). 
Our frame choice enables us to identify our geometric variables with physical quantities:  
\begin{itemize}
\item
$K_{ab}\equiv K_{(ab)}$ are the components of the extrinsic 3-curvature of ${\cal T}_t$, and 
\item $N_{ab}$ are the nine (intrinsic) spatial curvature variables.
\end{itemize}
Note that all Ricci rotation coefficients, $K_{ab}, N_{ab}$ act as scalars  on ${\cal T}_t$.
Here and throughout, parentheses denote symmetrization, {\it i.e.}, $K_{(ab)}\equiv {\textstyle \frac12} (K_{ab}+K_{ba})$.

By this frame choice, the tetrad Einstein-scalar evolution equations take the following form:
\begin{alignat}{2}
\label{eq-K-ab0}
& D_0 K_{ab} &&= \epsilon_{a}{}^{cd}D_c N_{db} + D_a b_b+ b_a b_b - \epsilon_{b}{}^{cd}N_{ac}b_d
\\ 
& &&+ N_c{}^cN_{ab}    
 - N_{ca}N^c{}_{b} + {\textstyle \frac12} \epsilon_{a}{}^{df}\epsilon_{b}{}^{ce}\left( K_{dc}K_{fe} - N_{dc}N_{fe} \right)
\nonumber\\
& &&- K_{a}{}^{c}K_{cb} + S_aS_b- {\textstyle \frac12}\delta_{ab}\big(W^2 + S^cS_c\big),
\nonumber\\
\label{eq-N-ab}
&D_0 N_{ab} &&= - \epsilon_{a}{}^{cd}D_c K_{db} 
+  \epsilon_{b}{}^{cd}K_{ac}b_d - N_c{}^c K_{ab} 
\\
& &&+ 2 N_{c[a}K_{b]}{}^c 
+  \epsilon_{a}{}^{df}\epsilon_{b}{}^{ce}N_{dc}K_{fe}
+ \epsilon_{ab}{}^{c}W S_c,
\nonumber\\
\label{eq-phi}
&D_0 \phi &&= W,\\
\label{eq-W}
&D_0 W &&= -\delta^{ab}K_{ab}W + D_a S^a + \left(b_a - 2A_a \right) S^a- V,_{\phi},\\
\label{eq-S-a0}
&D_0 S_a &&= D_a W + b_a W - K_{(ab)}S^b,
\end{alignat}
where $D_0$ is the Lie derivative along ${\bf e}_0$ and $D_a$ denotes the directional derivative along ${\bf e}_a$. Here, to separate out the antisymmetric part of $N_{ab}$, we introduced the new variable
\begin{equation}
A_b \equiv {\textstyle \frac12} \epsilon_{b}{}^{cd}N_{cd}.
\end{equation}

The evolution equations~(\ref{eq-K-ab0}-\ref{eq-S-a0}) are subject to the following constraints:
\begin{align}
\label{H-const}
 2\tensor{D}{_b} \tensor{A}{^b} &= {\textstyle \frac12} \tensor{K}{_a_b}\tensor{K}{^a^b} 
 - {\textstyle \frac12}(\tensor{K}{_a^a})^2 
 - {\textstyle \frac12}(\tensor{N}{_a^a})^2
 \\
&+ {\textstyle \frac12}\tensor{N}{_a_b}\tensor{N}{^a^b} + {\textstyle \frac12}W^2 + {\textstyle \frac12}\tensor{S}{^a}\tensor{S}{_a} + V(\phi),
\nonumber\\
\label{m-const}
\tensor{D}{_b} \tensor{K}{_a^b} - \tensor{D}{_a} \tensor{K}{_c^c} &= \tensor{\epsilon}{_a^b^c}\tensor{K}{_b^d} \tensor{N}{_d_c} + 2 \tensor{K}{_a^c} \tensor{A}{_c} + W \tensor{S}{_a},
\\
\label{N-ab-const}
\tensor{D}{_b} \tensor{N}{_a^b} - \tensor{D}{_a} \tensor{N}{_c^c} &= - \tensor{\epsilon}{_a^b^c}\tensor{N}{_b^d} \tensor{N}{_d_c} ,
\\
\label{const-phi}
\tensor{D}{_a} \phi &= \tensor{S}{_a}.
\end{align}

\subsection{Fixing the coordinate gauge}

To evolve the tetrad equations  numerically, we write them as a system of partial differential equations, {\it i.e.}, we re-express
directional derivatives along tetrads as partial derivatives along coordinate directions:
 \begin{equation}
 \label{def_partials}
D_0 = N^{-1}\left( \partial_t - N^i\partial_i \right),\quad D_a = E_a{}^i \partial_i,
\end{equation}
where $N<0$ is the tetrad lapse function; $t$ is running from zero to $+\infty$; $N^i$ are the three coordinate components of the tetrad shift vector; and $E_a{}^i$ describe projections of the spatial triad $\{\tensor{\bf e}{_a^\mu}\}$ tangent to the constant-time hypersurface ${\cal T}_t$. (Note that in the case of contracting spactimes as studied, {\it e.g.}, in Refs.~\cite{Ijjas:2020dws,Ijjas:2021gkf,Ijjas:2021wml}, $N>0$ but $t$ is running from zero to $-\infty$.)

The lapse and the components of the shift are gauge variables. We choose the coordinate gauge such that
\begin{itemize}
\item the coordinates are {\it co-moving} with the tetrad congruence ($N^i \equiv0$);
\item hypersurfaces of constant time ${\cal T}_t$ are {\it constant mean curvature} (CMC) hypersurfaces with the mean curvature $K\equiv \tensor{K}{_a^a}$ given by 
\begin{equation}
\label{def-time-app}
 -\frac{{\rm d} \ln K}{{\rm d}t} = \mu(t) ,
\end{equation} 
where $\mu(t)>0$. 
In the homogeneous and isotropic limit, $K=-3H$. 
\end{itemize}
The CMC gauge condition yields an elliptic equation for the coordinate lapse function:
\begin{align}
\label{lapse-eq}
K\partial_t (\ln K)
&=\Big(D_a - 2A_a\Big)D^aN \\
&- N\Big( \Sigma_{ab}\Sigma^{ab}
+ {\textstyle \frac13} K^2 + W^2 - V(\phi)\Big)
,\nonumber
\end{align}
where we introduced the new variable
\begin{equation}
\tensor{\Sigma}{_a_b} = \tensor{K}{_a_b} -{\textstyle \frac13} K\tensor{\delta}{_a_b}
\end{equation}
to denote the symmetric, trace-free part of the extrinsic 3-curvature $\tensor{K}{_a_b}$.
Note that the coordinates are also co-moving with the foliation $\{{\cal T}_t\}$ since the frame is hypersurface orthogonal.

\subsection{Evolution and constraint equations}

Under these frame and coordinate gauge conditions, 
the Einstein-scalar evolution equations take the following form:
\begin{alignat}{2}
 \label{eq-E-ai}
&\tensor{\partial}{_t} \tensor{E}{_a^i} 
&&= - N\Big( \tensor{\Sigma}{_a^c} 
+{\textstyle \frac13} K \tensor{\delta}{_a^c} \Big) \tensor{E}{_c^i} 
,\\
\label{eq-Sigma}
&\tensor{\partial}{_t} \tensor{\Sigma}{_a_b} 
&&= 
 N\Big( - K\tensor{\Sigma}{_a_b}+ \tensor{n}{_c^c}\tensor{n}{_\langle_a_b_\rangle}    
- 2 \tensor{n}{_\langle_a^c}\tensor{n}{_b_\rangle_c}
\Big)
\\
& &&+ N\Big(\tensor{\epsilon}{^c^d_\langle_a}\tensor{D}{_c} \tensor{n}{_b_\rangle_d}
-2\tensor{\epsilon}{^c^d_\langle_a}\tensor{A}{_c}\tensor{n}{_b_\rangle_d}
- \tensor{D}{_\langle_a}\tensor{A}{_b_\rangle}
+ \tensor{S}{_\langle_a}\tensor{S}{_b_\rangle}\Big)
\nonumber\\
& &&+  \tensor{D}{_\langle _a}\tensor{D}{_b_\rangle} N 
+\tensor{\epsilon}{^c^d_\langle_a}\tensor{n}{_b_\rangle_d} \tensor{D}{_c}N
+ \tensor{A}{_\langle_a}\tensor{D}{_b_\rangle}N 
,\nonumber\\
\label{eq-N-ab}
&\tensor{\partial}{_t} \tensor{n}{_a_b} 
&&= - N\Big({\textstyle \frac13} K\tensor{n}{_a_b}
- 2 \tensor{n}{_(_a^c} \tensor{\Sigma}{_b_)_c} 
+ \tensor{\epsilon}{_(_a^c^d}\tensor{D}{_c} \tensor{\Sigma}{_b_)_d} 
\Big)
\\
& &&- \tensor{\epsilon}{^c^d_(_a}\tensor{\Sigma}{_b_)_c}\tensor{D}{_d}N 
,
\nonumber\\
\label{eq-A-b}
&\tensor{\partial}{_t} \tensor{A}{_c} 
&&= - N\Big({\textstyle \frac13} K\tensor{A}{_c}
+ \tensor{\Sigma}{_c^b}\tensor{A}{_b}
-{\textstyle \frac12}\tensor{D}{_a} \tensor{\Sigma}{_c^a} 
\Big)
-{\textstyle \frac13} K\tensor{D}{_c}N
\\
& &&+ {\textstyle \frac12}\tensor{\Sigma}{^a_c}\tensor{D}{_a}N
,\nonumber\\
\label{eq-phi}
&\tensor{\partial}{_t} \phi &&= NW
,\\
\label{eq-W}
&\tensor{\partial}{_t} W &&= -N\Big(KW - \tensor{D}{_a} \tensor{S}{^a}  + 2\tensor{A}{_a} \tensor{S}{^a} + V,_{\phi}\Big) + \tensor{S}{^a}\tensor{D}{_a}N
,\\
\label{eq-S-a}
&\tensor{\partial}{_t} \tensor{S}{_a} &&= -N\big({\textstyle \frac13} K\tensor{S}{_a} + \tensor{\Sigma}{_a^b}\tensor{S}{_b}  - \tensor{D}{_a} W\Big) + W\tensor{D}{_a}N  
,
\end{alignat}
where we introduced the new variable
\begin{equation}
\tensor{n}{_a_b} = \tensor{N}{_a_b} - \tensor{\epsilon}{_a_b^c} \tensor{A}{_c},
\end{equation}
to denote the symmetric part of the intrinsic 3-curvature $\tensor{N}{_a_b}$ and eliminated the acceleration (gauge) vector $b_a$ using the relation $b_aN^{-1}= b_aD_0x^0=\left[{\bf e}_0, {\bf e}_a\right]x^0 = N^{-1}D_aN$.

The system is subject to the following constraint equations:
\begin{align}
\label{const-E-ai}
\tensor{\epsilon}{^b^c_a}\tensor{D}{_b} \tensor{E}{_c^i} 
&= \tensor{\epsilon}{^b^c_a} \tensor{A}{_b} \tensor{E}{_c^i} 
+ \tensor{n}{_a^d} \tensor{E}{_d^i},
\\
2D_b A^b &= -{\textstyle \frac13} K^2
 +{\textstyle \frac12}\tensor{\Sigma}{^a^b}\tensor{\Sigma}{_a_b} 
 + {\textstyle \frac12} \tensor{n}{^a^b}\tensor{n}{_a_b} 
  - {\textstyle \frac14}  (\tensor{n}{_a^a})^2 
 \\
& +3 \tensor{A}{^b}\tensor{A}{_b} + {\textstyle \frac12}W^2 + {\textstyle \frac12}S^aS_a + V(\phi)
,\nonumber\\
\label{m-const}
\tensor{D}{_b} \tensor{\Sigma}{_a^b}  
&= 3\tensor{\Sigma}{_a^b}\tensor{A}{_b}
+ \tensor{\epsilon}{_a^b^c}\tensor{n}{_b^d}\tensor{\Sigma}{_c_d}
+ W S_a
,\\
\label{N-ab-const}
D_b \tensor{n}{^b_a} 
&= - \tensor{\epsilon}{^b^c_a}\tensor{D}{_b} \tensor{A}{_c} + 2\tensor{A}{_b}\tensor{n}{^b_a}
,\\
\label{const-phi}
D_a \phi &= S_a.
\end{align}

\subsection{Re-scaling the variables}

Upon using the remaining gauge freedom to fix the time coordinate  $t$ as in Eq.~\eqref{newtime2}, we re-scale all variables as follows:
\begin{align}
\label{cal-N-def}
N &\rightarrow \tilde{\cal N} \equiv  {\textstyle \frac13}KN/\mu
,\\
 \partial_t &\rightarrow  \tilde{\partial}_t  \equiv \partial_t /\mu
,\\
E_a{}^i &\rightarrow \bar{E}_a{}^i\equiv E_a{}^i /\left({\textstyle \frac13} K\right)
,\\
\{\Sigma_{ab}, n_{ab}, A_b\} &\rightarrow  \{ \bar{\Sigma}_{ab} , \bar{n}_{ab}, \bar{A}_b  \} 
\equiv \{ \Sigma_{ab}, n_{ab}, A_b\}/\left({\textstyle \frac13} K\right), 
 \\
 \{W,  S_a\} &\rightarrow  \{\bar{W},  \bar{S}_a  \} 
\equiv \{W, S_a\} /\left({\textstyle \frac13} K\right),
 \\
\label{Vbar-def}
V &\rightarrow \bar{V} \equiv V/ \left({\textstyle \frac19} K^2\right).
\end{align}
This means, all geometric and scalar field variables are re-scaled by the mean curvature $K/3$, but the coordinate lapse $N$ is rescaled by the mean curvature and by $\mu$. The partial time derivative operator is re-scaled only by $\mu$, ensuring that the Lie derivative operator ($D_0$) in Eq.~\eqref{def_partials} is rescaled only by the mean curvature as are the directional derivatives along the spatial tetrads ($D_i$).

Substituting the rescaled variables into Eqs.~(\ref{lapse-eq}, \ref{eq-E-ai}-\ref{eq-S-a}), we obtain the elliptic lapse equation 
\begin{equation}
\label{lapse-eq-Hmu}
 -\Big(\bar{D}_a - 2\bar{A}_a\Big)\bar{D}^a\tilde{\cal N}
+ \Big( \bar{\Sigma}_{ab}\bar{\Sigma}^{ab}
+3  + \bar{W}^2 - \bar{V}(\phi)\Big)\tilde{\cal N} = 3
,
\end{equation}
and the Einstein-scalar evolution system takes the following form:
\begin{alignat}{2}
 \label{eq-E-ai-Hmu}
&\tilde{\partial}_t \bar{E}_a{}^i 
&&=  - \Big(\tilde{\cal N}-1\Big)\bar{E}_a{}^i 
- \tilde{\cal N}\bar{\Sigma}_a{}^c \bar{E}_c{}^i
,\\
\label{eq-Sigma-Hmu}
&\tilde{\partial}_t \tensor{\bar{\Sigma}}{_a_b} 
&&=- \Big(3\tilde{\cal N}-1\Big)\tensor{\bar{\Sigma}}{_a_b} 
 + \tilde{\cal N}\tensor{\bar{S}}{_\langle_a}\tensor{\bar{S}}{_b_\rangle}
 \\
& &&+ \tilde{\cal N}\Big( \tensor{\bar{n}}{_c^c}\tensor{\bar{n}}{_\langle_a_b_\rangle}    
- 2 \tensor{\bar{n}}{_\langle_a^c}\tensor{\bar{n}}{_b_\rangle_c}
+\tensor{\epsilon}{^c^d_\langle_a}\tensor{\bar{D}}{_c} \tensor{\bar{n}}{_b_\rangle_d}\Big)
\nonumber\\
& &&- \tilde{\cal N}\Big( 2\tensor{\epsilon}{^c^d_\langle_a}\tensor{\bar{A}}{_c}\tensor{\bar{n}}{_b_\rangle_d}
+ \tensor{\bar{D}}{_\langle_a}\tensor{\bar{A}}{_b_\rangle}\Big)
\nonumber\\
& &&+  \tensor{\bar{D}}{_\langle _a}\tensor{\bar{D}}{_b_\rangle} \tilde{\cal N} 
+\tensor{\epsilon}{^c^d_\langle_a}\tensor{\bar{n}}{_b_\rangle_d} \tensor{\bar{D}}{_c}\tilde{\cal N}
+ \tensor{\bar{A}}{_\langle_a}\tensor{\bar{D}}{_b_\rangle}\tilde{\cal N} 
,\nonumber\\
\label{eq-N-ab-Hmu}
&\tilde{\partial}_t \tensor{\bar{n}}{_a_b} 
&&= - \Big(\tilde{\cal N}-1\Big)\tensor{\bar{n}}{_a_b} 
+\tilde{\cal N}\Big(
 2 \tensor{\bar{n}}{_(_a^c} \tensor{\bar{\Sigma}}{_b_)_c} 
- \tensor{\epsilon}{_(_a^c^d}\tensor{\bar{D}}{_c} \tensor{\bar{\Sigma}}{_b_)_d} 
\Big)
\\
& &&- \tensor{\epsilon}{^c^d_(_a}\tensor{\bar{\Sigma}}{_b_)_c}\tensor{\bar{D}}{_d}\tilde{\cal N} 
,
\nonumber\\
\label{eq-A-b-Hmu}
&\tilde{\partial}_t \tensor{\bar{A}}{_c} 
&&= - \Big(\tilde{\cal N}-1\Big)\tensor{\bar{A}}{_c} 
- \tilde{\cal N}\Big( \tensor{\bar{\Sigma}}{_c^b}\tensor{\bar{A}}{_b}
-{\textstyle \frac12}\tensor{\bar{D}}{_a} \tensor{\bar{\Sigma}}{_c^a} 
\Big)
-\tensor{\bar{D}}{_c}\tilde{\cal N}
\\
& &&+ {\textstyle \frac12}\tensor{\bar{\Sigma}}{^a_c}\tensor{\bar{D}}{_a}\tilde{\cal N}
,\nonumber\\
\label{eq-phi-Hmu}
&\tilde{\partial}_t \phi 
&&= \tilde{\cal N}\bar{W}
,\\
\label{eq-W-Hmu}
&\tilde{\partial}_t \bar{W} 
&&= - \Big(3\tilde{\cal N}-1\Big)\bar{W}  - \tilde{\cal N}\Big( \bar{V},_{\phi}- \bar{D}_a \bar{S}^a  + 2\bar{A}_a \bar{S}^a \Big) \\
& &&+ \bar{S}^a\bar{D}_a\tilde{\cal N}
,\nonumber\\
\label{eq-S-a-Hmu}
&\tilde{\partial}_t \bar{S}_a 
&&= \bar{S}_a -\tilde{\cal N}\big(\bar{S}_a + \tensor{\bar{\Sigma}}{_a^b}\bar{S}_b  - \bar{D}_a \bar{W}\Big) + \bar{W}\bar{D}_a\tilde{\cal N}  
.
\end{alignat}

The evolution system is subject to the rescaled constraint equations:
\begin{align}
\label{hamiltonian-const}
 2\bar{D}_b \bar{A}^b &= -3
 +{\textstyle \frac12}\tensor{\bar{\Sigma}}{^a^b}\tensor{\bar{\Sigma}}{_a_b} 
 + {\textstyle \frac12} \tensor{\bar{n}}{^a^b}\tensor{\bar{n}}{_a_b} 
  - {\textstyle \frac14}  (\tensor{\bar{n}}{_a^a})^2 
  \\
& +3 \tensor{\bar{A}}{^b}\tensor{\bar{A}}{_b}
+ {\textstyle \frac12}\bar{W}^2 + {\textstyle \frac12}\bar{S}^a\bar{S}_a + \bar{V}(\phi)
,\nonumber\\
\label{m-const}
\tensor{D}{_b} \tensor{\bar{\Sigma}}{_a^b}  &= 
3\tensor{\bar{\Sigma}}{_a^b}\tensor{\bar{A}}{_b}
+ \tensor{\epsilon}{_a^b^c}\tensor{\bar{n}}{_b^d}\tensor{\bar{\Sigma}}{_c_d}
+ \bar{W} \bar{S}_a
,\\
\label{N-ab-const}
\bar{D}_b \tensor{\bar{n}}{^b_a} &= - \tensor{\epsilon}{^b^c_a}\tensor{\bar{D}}{_b} \tensor{\bar{A}}{_c} + 2\tensor{\bar{A}}{_b}\tensor{\bar{n}}{^b_a}
,\\
\label{const-E-ai-hn}
\tensor{\epsilon}{^b^c_a}\tensor{\bar{D}}{_b} \tensor{\bar{E}}{_c^i} &= \tensor{\epsilon}{^b^c_a} \tensor{\bar{A}}{_b} \tensor{\bar{E}}{_c^i} 
+ \tensor{\bar{n}}{_a^d} \tensor{\bar{E}}{_d^i},
\\
\label{const-phi-Hmu}
\bar{D}_a \phi &= \bar{S}_a.
\end{align}

\section{Homogeneous limit}
\label{app:homogeneous}

In the homogeneous limit ($\tensor{E}{_a^i}, \tensor{A}{_b}, \tensor{S}{_a}\rightarrow0$), the lapse equation~\eqref{lapse-eq-Hmu} becomes a simple algebraic constraint:
\begin{equation}
\label{lapse-eq-hom}
3 =  \tilde{\cal N}\Big( \bar{\Sigma}_{ab}\bar{\Sigma}^{ab}
+3  + \bar{W}^2 - \bar{V}(\phi)\Big)
.
\end{equation}

The evolution equations~(\ref{eq-E-ai-Hmu}, \ref{eq-A-b-Hmu}) and \eqref{eq-S-a-Hmu} for $\tensor{\bar{E}}{_a^i}, \tensor{\bar{A}}{_b}, \tensor{\bar{S}}{_a}$, respectively, are trivially satisfied and the remaining equations~(\ref{eq-Sigma-Hmu}-\ref{eq-N-ab-Hmu}, \ref{eq-phi-Hmu}-\ref{eq-W-Hmu}) reduce to the simple system of ordinary differential equations:
\begin{alignat}{2}
\label{eq-Sigma-hom}
&\tilde{\partial}_t \tensor{\bar{\Sigma}}{_a_b} 
&&= - \Big(3\tilde{\cal N}-1\Big) \tensor{\bar{\Sigma}}{_a_b} 
+ \tilde{\cal N}\Big( \tensor{\bar{n}}{_c^c}\tensor{\bar{n}}{_\langle_a_b_\rangle}    
- 2 \tensor{\bar{n}}{_\langle_a^c}\tensor{\bar{n}}{_b_\rangle_c}
\Big),\\
\label{eq-N-ab-hom}
&\tilde{\partial_t} \tensor{\bar{n}}{_a_b} 
&&= - \Big(\tilde{\cal N}-1\Big)\tensor{\bar{n}}{_a_b}  + 2 \tilde{\cal N}\tensor{\bar{n}}{_(_a^c} \tensor{\bar{\Sigma}}{_b_)_c} ,\\
\label{eq-phi-hom}
&\tilde{\partial}_t \phi &&= \tilde{\cal N}\bar{W},\\
\label{eq-W-hom}
&\tilde{\partial}_t \bar{W}  
&&= -\Big(3\tilde{\cal N}-1\Big)\bar{W}  -\tilde{\cal N} \bar{V},_{\phi},
\end{alignat}
that is supplemented by the homogeneous Hamiltonian and momentum constraints:
\begin{align}
& {\textstyle \frac12}\tensor{\bar{\Sigma}}{^a^b}\tensor{\bar{\Sigma}}{_a_b} 
 + {\textstyle \frac12} \tensor{\bar{n}}{^a^b}\tensor{\bar{n}}{_a_b} 
  - {\textstyle \frac14}  (\tensor{\bar{n}}{_a^a})^2 
+ {\textstyle \frac12}\bar{W}^2 + \bar{V}(\phi) = 3
,\\
\label{m-const-hom}
 &  \tensor{\epsilon}{_a^b^c}\tensor{\bar{n}}{_b^d}\tensor{\bar{\Sigma}}{_c_d} = 0
.
\end{align}
This set of equations is equivalent to those discussed in Sec.5 of Ref.~\cite{Ijjas:2020dws}, upon exchanging $\{{\cal N}, \partial_t\}\leftrightarrow\{\tilde{\cal N}, \tilde\partial_t\}$. That means, the dynamical stability of the system is fully determined by the eigenvalue evolution and the critical points of the system correspond to those listed in Table 1 of Ref.~\cite{Ijjas:2020dws}.

In particular, in the limit of flat FRW spacetimes, the critical point solution is given by 
\begin{equation}
\label{FRWcriticalpoint}
\tilde{\cal N}^{-1} = {\textstyle \frac12}\left(\frac{\bar{V}_{\phi}}{\bar{V}}\right)^2, \quad 
\bar{W}^2 = \left(\frac{\bar{V}_{\phi}}{\bar{V}}\right)^2,\quad
\bar{V} = 3 - {\textstyle \frac12}\bar{W}^2.
\end{equation}
In an expanding homogeneous spacetime, the solution is a stable critical point if $\tilde{\cal N}^{-1}>1$, which is the inflationary  solution.

\section{Rescaled time coordinate}
\label{app:newtime}

Here, we demonstrate that our new  time coordinate $t$ as given in Eq.~\eqref{newtime2},
measures the {\it maximum} number of $e$-folds of inflation that can occur. 
As above, $K\equiv \tensor{K}{_a^a}$ is the trace of the extrinsic 3-curvature $K_{ab}$; and $\mu(t)$ is a positive definite, purely time dependent function. 

For an inflating patch, the effective equation of state $\varepsilon<1$ is (approximately) constant and the FRW scale factor $a(\tau)$ and Hubble radius $H^{-1}$ obey simple scaling relations,
\begin{equation}
\label{scaling-sol}
a(\tau)\simeq \tau^{1/\varepsilon},\quad 
H^{-1}\simeq \varepsilon \tau ,
\end{equation}
where  $\tau$ is the (physical) FRW time coordinate.
The number of  $e$-folds of inflation $N_{\rm aH}$ in the patch is given by
\begin{equation}
\label{def-efold}
N_{\rm aH} = \ln \left( \frac{a}{H^{-1}} \Big/ \frac{a_0}{H^{-1}_0}\right) \simeq 
  \left( \frac{1}{\varepsilon}-1\right) \ln  \left( \frac{\tau}{\tau_0}\right),
\end{equation}
where $a_0 = \tau_0^{1/\varepsilon}$, $H_0^{-1}=\varepsilon\tau_0$ and $\tau_0$ is the time denoting the beginning of inflation. For simplicity but without loss of generality, we set $\tau_0=1$, such that $a_0=1$ and $H_0=1/\varepsilon$. 

In the inflating patch, $-3/K = H^{-1}$ is the Hubble radius and  
\begin{equation}
\varepsilon \simeq (1/2)(V_{\phi}/V)^2,
\end{equation}
 see, {\it e.g.}, Ref.~\cite{Steinhardt:1984jj}. 

Using Eqs.~(\ref{newtime2}) and (\ref{scaling-sol}), it is straightforward to verify that in this patch the FRW time coordinate $\tau$ and the coordinate time $t$ are related by
\begin{equation}
\label{tautot}
\ln  \tau =  \int \mu(t) {\rm d}t.
\end{equation}
Furthermore, with Eq.~\eqref{FRWcriticalpoint}, we find 
\begin{equation}
\tilde{\cal N} =1/\varepsilon.
\end{equation}

For sufficiently small time intervals $\Delta t$, $\mu$ is approximately constant such that $t$ measures the number of $e$-folds of inflation $N_{\rm aH}$ if 
\begin{equation}
\mu =  \left( \tilde{\cal N}-1\right)^{-1}.
\end{equation}
Therefore, for $t$ to measure the maximum number of $e$-folds of inflation $N_{\rm aH}$ taking place anywhere within the simulated spacetime region, we choose 
\begin{equation}
\label{def-mu}
\mu(t) \equiv \frac{1}{\tilde{\cal N}_{\rm max}(t)}
,
\end{equation}
with ${\cal N}_{\rm max}\equiv {\rm max}({\cal N} \, |\, t = {\rm const.})$ denoting the maximum value of ${\cal N}$ at time $t$.

\section{Measure for proper volume}

Here we demonstrate that the mean-curvature-normalized lapse ${\cal N}$ is a natural measure of proper volume.

The proper spatial volume element $S $ of a patch with spatial 3-metric $\gamma_{ij}$ is given by 
\begin{equation}
S \equiv \sqrt{\gamma}.
\end{equation}
That is, the rate of change in proper volume is given by 
\begin{equation}
\label{proper-vol}
{\cal L}_{{\bf e}_0} \ln S =
- \frac12 N^{-1} \gamma^{ij}  \partial_t  \gamma_{ij} = 
K ,
\end{equation}
where ${\cal L}_{{\bf e}_0}\equiv N^{-1}\partial_t$ denotes the Lie derivative along the evolution normal vector ${\bf e}_0$.
In particular,
 \begin{equation}
\partial_t \ln S = NK 
= 3\mu(t)\tilde{\cal N},
 \end{equation}
such that we can use the mean-curvature-normalized lapse ${\cal N} \equiv {\textstyle \frac13} NK$, to compute the (proper) volume ratio between patches that inflate and those that do not.

The relation in Eq.~\eqref{proper-vol} becomes immediately apparent in the homogeneous limit. Here, $\tilde{\cal N} = 1/\varepsilon$ and the proper volume is given by  
\begin{equation}
\label{propervolume}
S = e^{\frac{3}{\varepsilon} \int \mu(t) {\rm d}t }
= \left( e^{\frac{\varepsilon_{\rm min}}{\varepsilon}\, t} \right)^3 ,
\end{equation}
{\it i.e.}, regions with $\varepsilon>\varepsilon_{\rm min}$ occupy exponentially less volume than regions with the smallest equation of state ($\varepsilon = \varepsilon_{\rm min}$).
Also note that, with Eq.~\eqref{tautot}, we can re-write Eq.~\eqref{propervolume} using the (physical) FRW time  coordinate and obtain $S = (\tau^{1/\varepsilon})^3 = a^3$, the known result.

\section{Initial conditions}
\label{app:init}

Here, we briefly explain how the initial data is being fixed using York's conformal method \cite{York:1971hw} and connect our initial data in Sec.~\ref{sec:formalism} expressed in (mean-curvature-normalized) variables to those used in Refs.~\cite{East:2015ggf,Clough:2016ymm,Clough:2017efm, Aurrekoetxea:2019fhr,Joana:2020rxm,Corman:2022alv}.

Specifying the initial conditions means fixing the 3-metric $\gamma_{ij}$ of a spacelike hypersurface ${\cal T}_{t_0}$ at some initial time $t_0$, its time derivative $D_0\gamma_{ij}$ as well as the initial energy and momentum densities, $\rho(t_0)$ and $p_i(t_0)$. Since we have the freedom to set the initial lapse $N$ to -1 and the initial shift $N_i$ to zero,  determining the time derivative of the spatial metric at $t_0$ is equivalent to determining the  extrinsic curvature $K_{ij}$ at $t_0$, $D_0\gamma_{ij} = (N^{-1}\partial_t + N^i\partial_i)\gamma_{ij} = -\partial_t\gamma_{ij} = 2K_{ij}$. 

However, we do not have complete freedom to choose the initial conditions. This is because, for an arbitrary combination of $\gamma_{ij}, K_{ij}, \rho$ and $p_j$, the Einstein equations need not satisfy the Hamiltonian and momentum constraints, 
\begin{alignat}{1}
\label{hc}
{}^{(3)}R + K^2 - K_{ij}K^{ij} &= 2\rho,\\
\label{mc}
 D^iK_{ij} - D_jK &= p_j
.
\end{alignat}
If we  choose constraint satisfying initial conditions, the field equations preserve and propagate the constraints such that those remain satisfied at all times.

In numerical relativity, it is common to employ York's method  to ensure that the chosen initial conditions satisfy the constraints~(\ref{hc}-\ref{mc}).
This method relies on conformally rescaling the spatial metric,
$\gamma_{ij}$,
and then rescaling the variables, $K_{ij}, \rho$ and $p_i$, by an appropriate power of the conformal factor $\psi$ such that the momentum constraint decouples from the Hamiltonian constraint. 

More precisely, we re-express $\gamma_{ij}$ and $K_{ij}$ as follows:
\begin{align}
\gamma_{ij} &= \psi^4(t_0, {\bf x})\tilde{\gamma}_{ij},\\
K_{ij} &= \psi^{-2}A_{ij}+ \frac13 K \psi^4\tilde{\gamma}_{ij}.
\end{align} 
Here, we additionally require that the spatial metric $\gamma_{ij}$ is conformally flat, {\it i.e.}, $\tilde{\gamma}_{ij}\equiv\delta_{ij}$, as also required in all inflationary studies to date.

Substituting into Eqs.~(\ref{hc}-\ref{mc}), the Hamiltonian and momentum constraints take the simple(r) form:
\begin{alignat}{2}
\label{hamiltonian-c}
& \partial_i\partial^i\psi &&+ {\textstyle \frac18}\psi^{-7} \big(\tensor{A}{^i^j}\tensor{A}{_i_j} + \eta^2 \big) 
- {\textstyle \frac{1}{12}}\psi^5 \big(K^2 - 3 V \big)
\\
& &&+ {\textstyle \frac18}\psi \partial^i\phi\partial_i\phi  = 0 ,
\nonumber\\
\label{momentum-c}
&\partial^i\tensor{A}{_i_j}  &&= -\eta\partial_j\phi,
\end{alignat}
where we used that, for a canonical scalar minimally coupled to Einstein gravity with potential $V(\phi)$,
\begin{align}
\label{rho-phi}
\rho &= {\textstyle \frac12} \dot{\phi}^2 
+ {\textstyle \frac12} D^i\phi D_i\phi + V(\phi)\\
&= {\textstyle \frac12}\psi^{-4} \left(\psi^{-8}\eta^2 + \partial^i\phi \partial_i\phi \right) + V(\phi)
,\nonumber\\
\label{p-phi}
p_j & = - \dot{\phi}D_j\phi = - \psi^{-6}\eta\partial_j\phi
,
\end{align}
with  
\begin{equation}
\eta \equiv \psi^6 \dot{\phi}
\end{equation}
being the conformally rescaled scalar field velocity,
dot denoting the Lie derivative along the timelike normal vector and $D_j$ denoting the covariant derivative w.r.t. $\gamma_{ij}$.

It is immediately apparent that 
the momentum constraint~\eqref{momentum-c} only depends on $A_{ij}$ and its solution is always of the form:
\begin{equation}
A_{ij} = A_{ij}^{TT} + A_{ij}^L ,
\end{equation}
where $A_{ij}^{TT}$ denotes the divergence-free vacuum shear contribution ($\partial^iA_{ij}^{TT} = 0$)
 and $A_{ij}^L= A_{ij} - A_{ij}^{TT}$ denotes the longitudinal part of the conformally rescaled trace-free shear $\tensor{A}{_i_j}$. Note that because $A_{ij}^{TT}$ is transverse and traceless, it is sometimes called the initial gravitational wave contribution. 

Accordingly, by selecting the initial conditions, we can freely specify 
\begin{equation}
\label{freedata}
\{K, A_{ij}^{TT}, \phi, \eta\}.
\end{equation}
Then, substituting $\phi$ and $Q$ into the momentum constraint~\eqref{momentum-c}, we obtain $A_{ij}^L$. Finally, substituting $K, \phi, \eta, A_{ij}^{TT}, A_{ij}^L$ and $V(\phi)$ into the Hamiltonian constraint, we numerically solve the elliptic equation~\eqref{hamiltonian-c} for the conformal factor $\psi$ and obtain a complete set of initial data.

We note that, introducing $\tilde{\eta}= \psi^{-4}\eta= \psi^2\dot{\phi}$, as suggested in Ref.~\cite{Corman:2022alv}, both the Hamiltonian and momentum constraint explicitly depend on both $\psi$ and $A_{ij}$. An analytic ansatz as used in Ref.~\cite{Corman:2022alv} represents a select, non-generic type of initial data and hence cannot be used to make any generally valid conclusion. 

Finally, the variables $K_0, \phi, \eta, A_{ij}^{TT}, A_{ij}^L, \psi$ translate to the variables in mean-curvature-normalized tetrad form as introduced above in Sec.~\ref{sec:formalism}:
\begin{align}
\{K_0, \phi, \psi \}&= \{K_0, \phi, \psi \},\\
\eta &= \bar{Q} \times \left({\textstyle \frac13}K\right), \\
A_{ij}^{TT} &= -\bar{Z}_{ab}^0 \times \left({\textstyle \frac13}K\right), \\
A_{ij}^L &= -(\bar{Z}_{ab}-\bar{Z}_{ab}^0)\times \left({\textstyle \frac13}K\right).
\end{align}

\end{document}